# A Criterion for Parameter Identification in Structural Equation Models

**Jin Tian**
Department of Computer Science
Iowa State University
Ames, IA 50011
*jtian@cs.iastate.edu*

## Abstract

This paper deals with the problem of identifying direct causal effects in recursive linear structural equation models. The paper establishes a sufficient criterion for identifying individual causal effects and provides a procedure computing identified causal effects in terms of observed covariance matrix.

## 1 Introduction

Structural equation models (SEMs) have dominated causal reasoning in the social sciences and economics, in which interactions among variables are usually assumed to be linear [Duncan, 1975, Bollen, 1989]. This paper deals with one fundamental problem in SEMs, accessing the strength of linear cause-effect relationships from a combination of observational data and model structures.

The problem has been under study for half a century, primarily by econometricians and social scientists, under the name "The Identification Problem"[Fisher, 1966]. Although many algebraic or graphical methods have been developed, the problem is still far from being solved. In other words, we do not have a necessary and sufficient criterion for deciding whether a causal effect can be computed from observed data. Most available methods are sufficient criteria which are applicable only when certain restricted conditions are met.

In this paper, we provide a new sufficient criterion for the identification of individual causal effects. Our method is based on the partial regression equations, which reduce the identification problem into a problem of solving a set of algebraic equations [Tian, 2005]. We present a procedure that will determine sufficient conditions under which these equations can be solved and express identified causal effects in terms of observed covariances.

We begin with an introduction to SEMs and the identification problem, and give a brief review to previous work before presenting our results. Due to space constraints, the proofs are not given. Proofs of all results are given in [Tian, 2007].

## 2 Linear SEMs and Identification

A linear SEM over a set of random variables $V = \{V_1, \ldots, V_n\}$ is given by a set of structural equations of the form

$$V_j = \sum_i c_{ji} V_i + \epsilon_j, \quad j = 1, \ldots, n, \qquad (1)$$

where the summation is over the variables in $V$ judged to be immediate causes of $V_j$. $c_{ji}$, called a *path coefficient*, quantifies the direct causal influence of $V_i$ on $V_j$, and is also called a *direct effect*. $\epsilon_j$'s represent "error" terms and are assumed to have normal distribution. In this paper we consider recursive models and assume that the summation in Eq. (1) is for $i < j$, that is, $c_{ji} = 0$ for $i \geq j$. The set of variables (and the corresponding structural equations) are considered to be ordered as $V_1 < V_2 < \ldots < V_n$. We denote the covariances between observed variables $\sigma_{ij} = Cov(V_i, V_j)$, and between error terms $\psi_{ij} = Cov(\epsilon_i, \epsilon_j)$. We denote the following matrices, $\Sigma = [\sigma_{ij}]$, $\Psi = [\psi_{ij}]$, and $C = [c_{ij}]$. Without loss of generality, the model is assumed to be standardized such that each variable $V_j$ has zero mean and variance 1.

The structural assumptions encoded in the model are the zero path coefficient $c_{ji}$'s and zero error covariance $\psi_{ij}$'s. The model structure can be represented by a directed acyclic graph (DAG) $G$ with (dashed) bidirected edges, called a *causal diagram* (or *path diagram*), as follows: the nodes of $G$ are the variables $V_1, \ldots, V_n$; there is a directed edge from $V_i$ to $V_j$ in $G$ if $V_i$ appears in the structural equation for $V_j$, that is, $c_{ji} \neq 0$; there is a bidirected edge between $V_i$ and $V_j$ if the error terms $\epsilon_i$ and $\epsilon_j$ have non-zero correlation ($\psi_{ij} \neq 0$). Figure 1 shows a simple SEM and the corresponding causal diagram in which each directed edge is annotated by the corresponding path coefficient.

The parameters of the model are the non-zero entries in the

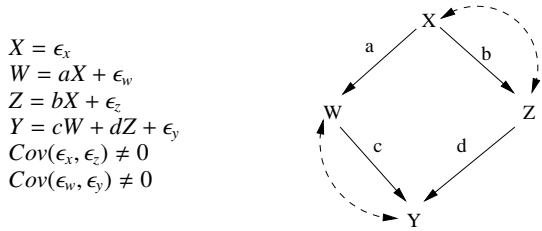

$X = \epsilon_x$
$W = aX + \epsilon_w$
$Z = bX + \epsilon_z$
$Y = cW + dZ + \epsilon_y$
$Cov(\epsilon_x, \epsilon_z) \neq 0$
$Cov(\epsilon_w, \epsilon_y) \neq 0$

Figure 1: A linear SEM.

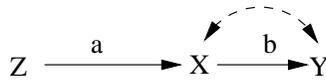

Figure 2: A typical instrumental variable

matrices $C$ and $\Psi$. Fixing the model structure and given parameters $C$ and $\Psi$, the covariance matrix $\Sigma$ is given by (see, for example, [Bollen, 1989])

$$\Sigma = (I - C)^{-1}\Psi(I - C)^{t-1}. \qquad (2)$$

Conversely, in the identification problem, given the structure of a model, one attempts to solve for $C$ in terms of the given observed covariance matrix $\Sigma$. If Eq. (2) gives a unique solution to some path coefficient $c_{ji}$, independent of the (unobserved) error correlations $\Psi$, the path coefficient $c_{ji}$ is said to be *identified*; otherwise, $c_{ji}$ is said to be *non-identifiable*. In other words, the *identification problem* is that whether a path coefficient is determined uniquely from the covariance matrix $\Sigma$ given the causal diagram. If every parameter of the model is identified, then *the model is identified*. Note that the identifiability conditions we seek involve the structure of the model alone, not particular numerical values of parameters, that is, we insist on having *identification almost everywhere*, allowing for pathological exceptions (see, for example, [Brito and Pearl, 2002a] for formal definition of identification almost everywhere).

## 3 Previous Work

Many methods have been developed for deciding whether a specific parameter or a model is identifiable. For example, the well-known instrumental variable (IV) method [Bowden and Turkington, 1984] requires searching for variables (called *instruments*) that are uncorrelated with the error terms in specific equations. A typical configuration of the IV method is shown in Fig. 2, in which $Z$ serves as an instrument for identifying the causal effect $b$ as

$$b = \sigma_{ZY}/\sigma_{ZX}. \qquad (3)$$

Traditional approaches are based on algebraic manipulation of the structural equations [Fisher, 1966, Bekker et al., 1994, Rigdon, 1995]. In recent years graphical approaches for identifying linear causal effects have been developed, and some sufficient graphical conditions for identification were established [McDonald, 1997, Pearl, 1998, Spirtes et al., 1998, Pearl, 2000, Spirtes et al., 2001, Tian, 2004]. The applications of these methods are limited in scope, and typically some special conditions have to be met for these methods to be applicable.

One principled approach for the identification problem is to write Eq.(2) for each term $\sigma_{ij}$ of $\Sigma$ using Wright's method of path coefficients [Wright, 1934]. Wright's equations consist of equating the (standardized) covariance $\sigma_{ij}$ with the sum of products of parameters ($c_{kl}$'s and $\psi_{kl}$'s) along certain paths between $V_i$ and $V_j$ in the causal diagram. For example, the Wright's equations for the model in Fig. 2 are

$$\sigma_{ZX} = a \qquad (4)$$
$$\sigma_{ZY} = ab \qquad (5)$$
$$\sigma_{XY} = b + \psi_{XY}. \qquad (6)$$

Then a path coefficient $c_{ij}$ is identified if and only if Wright's equations give a unique solution to $c_{ij}$, independent of error correlation $\psi_{kl}$'s.

Based on Wright's equations, a number of sufficient graphical criteria for identification have been developed. [Brito and Pearl, 2002b, Brito and Pearl, 2006] defined "auxiliary set" to obtain a set of linearly independent equations with path coefficients $c_{jk}$'s as unknowns. However, the coefficients of these equations are functions of other parameters which must be identified before solving the equations for $c_{jk}$'s. So they established sufficient criteria for *model identification*, that is, conditions for *every* parameter in the model to be identified. [Brito and Pearl, 2002a] defined "instrumental set" also obtaining a set of linearly independent equations with path coefficients as unknowns. The approach provides sufficient conditions for identifying individual path coefficients. The main difficulty seems to be that it is not an easy task to find out an instrumental set.

Recently, another principled approach for the identification problem is presented in [Tian, 2005], in which the partial regression coefficients, instead of the covariance, are expressed in terms of model parameters obtaining so called partial regression equations. Then a path coefficient $c_{ij}$ is identified if and only if the set of partial regression equations give a unique solution to $c_{ij}$, independent of $\psi_{kl}$'s.

In this paper, we will derive sufficient conditions for identifying individual path coefficients based on partial regression equations. We acknowledge that some of the ideas in this paper were inspired by the work in [Brito and Pearl, 2002c, Brito and Pearl, 2002b, Brito and Pearl, 2002a, Brito and Pearl, 2006]. First we introduce the partial regression equations.

## 4 Partial Regression Equations

For a set $S \subseteq V$, let $\beta_{ij.S}$ denote the *partial regression coefficient* which represents the coefficient of $V_j$ in the linear regression of $V_i$ on $V_j$ and $S$. (Note that the order of the subscripts in $\beta_{ij.S}$ is essential.) Partial regression coefficients can be expressed in terms of covariance matrices as follows [Cramer, 1946]:

$$\beta_{ij.S} = \frac{\Sigma_{V_i V_j} - \Sigma_{V_i S}^T \Sigma_{SS}^{-1} \Sigma_{V_j S}}{\Sigma_{V_j V_j} - \Sigma_{V_j S}^T \Sigma_{SS}^{-1} \Sigma_{V_j S}}, \quad (7)$$

where $\Sigma_{SS}$ etc. represent covariance matrices over corresponding variables.

Let $S_{jk}$ denote a set

$$S_{jk} = \{V_1, \ldots, V_{j-1}\} \setminus \{V_k\}. \quad (8)$$

[Tian, 2005] derived an expression for the partial regression coefficient $\beta_{jk.S_{jk}}$ in terms of the model parameters (path coefficients and error covariances) given by

$$\beta_{jk.S_{jk}} = c_{jk} + \alpha_{jk} - \sum_{k+1 \leq l \leq j-1} \beta_{lk.S_{lk}} \alpha_{jl},$$
$$j = 2, \ldots, n, \ k = 1, \ldots, j-1, \quad (9)$$

where $\alpha_{jk}$'s are defined during the process of "orthogonalizing" the set of error terms to obtain a new set of error terms $\{\epsilon_1', \ldots, \epsilon_n'\}$ that are mutually orthogonal in the sense that

$$Cov(\epsilon_i', \epsilon_j') = 0, \ \text{for } i \neq j. \quad (10)$$

The Gram-Schmidt orthogonalization process proceeds recursively as follows. We set

$$\epsilon_1' = \epsilon_1. \quad (11)$$

For $j = 2, \ldots, n$, we set

$$\epsilon_j' = \epsilon_j - \sum_{k=1}^{j-1} \alpha_{jk} \epsilon_k', \quad (12)$$

in which

$$\alpha_{jk} = \frac{Cov(\epsilon_j, \epsilon_k')}{Cov(\epsilon_k', \epsilon_k')}. \quad (13)$$

The set of equations given by (9) are called the *partial regression equations*. As an example, the partial regression equations for the model shown in Figure 1 are given by

$$\beta_{WX} = a \quad (14)$$
$$\beta_{ZW.X} = 0 \quad (15)$$
$$\beta_{ZX.W} = b + \alpha_{ZX} \quad (16)$$
$$\beta_{YZ.WX} = d \quad (17)$$
$$\beta_{YW.XZ} = c + \alpha_{YW} \quad (18)$$
$$\beta_{YX.WZ} = -\beta_{WX} \alpha_{YW} \quad (19)$$

It is not difficult to solve these equations to obtain that the path coefficients $a$, $d$, and $c$ are identified.

In general, given the model structure (represented by zero path coefficients and zero error correlations), some of the $c_{jk}$'s and $\alpha_{jk}$'s will be set to zero in Eq. (9), and we can solve the identifiability problem by solving Eq. (9) for $c_{jk}$'s in terms of the partial regression coefficients. This provides a principled method for solving the identifiability problem. A path coefficient $c_{ij}$ is identified if and only if the set of partial regression equations give a unique solution to $c_{ij}$, independent of error correlations.

The partial regression equations are linear with respect to path coefficient $c_{jk}$'s and parameter $\alpha_{jk}$'s, but may not be linear with respect to $\psi_{ij}$'s. $\alpha_{jk}$'s are nonlinear functions of $\psi_{ij}$'s and may not be independent with each other. However, we can treat $\alpha_{jk}$'s as independent unknowns and solve the partial regression equations (9). If we find a unique solution to $c_{jk}$, then $c_{jk}$ is identified. Although we can not say $c_{jk}$ is nonidentifiable if we could not find a unique solution to $c_{jk}$. In this paper, we will treat $\alpha_{jk}$'s as independent unknowns and look for sufficient conditions for identifying path coefficients. The identification problem is reduced to that of solving the set of linear equations (9) for $c_{jk}$'s in terms of the partial regression coefficients $\beta_{jk.S_{jk}}$'s, and a path coefficient $c_{jk}$ is identified if the set of equations give a unique solution to $c_{jk}$ that is independent of $\alpha_{jk}$'s.

## 5 Identifying Path Coefficients

Assume that we want to identify the path coefficients associated with a variable $V_j$, $c_{jk}$'s. All we need to do is to solve the $j-1$ equations given in (9) for $k = 1, \ldots, j-1$. We will name each of the equations in (9) after the corresponding variable as in the following:

$$(V_k) : \beta_{jk.S_{jk}} = c_{jk} + \alpha_{jk} - \sum_{k+1 \leq l \leq j-1} \beta_{lk.S_{lk}} \alpha_{jl}. \quad (20)$$

Let $V_j^< = \{V_1, \ldots, V_{j-1}\}$ denote the set of variables ordered ahead of $V_j$. Let $Pa_j$ be the set of variables $V_k$ such that $c_{jk} \neq 0$, that is, there is a directed edge from $V_k \rightarrow V_j$ in the causal diagram. Let $NP_j = V_j^< \setminus Pa_j$. $\alpha_{jk}$'s are nonlinear functions of $\psi_{ij}$'s and some of $\alpha_{jk}$'s will be identically zero depending on the model structure. Let $Ne_j$ be the set of variables $V_k$ such that $\alpha_{jk}$ is not identically zero denoted by $\alpha_{jk} \not\equiv 0$. In summary, we classify the set of variables in $V_j^<$ into possibly overlapping groups:

$$Pa_j = \{V_k | V_k \in V_j^<, c_{jk} \neq 0\}. \quad (21)$$
$$NP_j = \{V_k | V_k \in V_j^<, c_{jk} = 0\}. \quad (22)$$
$$Ne_j = \{V_k | V_k \in V_j^<, \alpha_{jk} \not\equiv 0\}. \quad (23)$$

Let $E(S)$ denote the set of equations $(V_k)$ such that $V_k \in S$. Each equation $(V_k)$ in $E(Pa_j)$ can be solved for the path

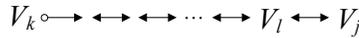

Figure 3: An active path between $V_k$ and $V_j$ given $S_{jk}$

coefficient $c_{jk}$ by simply rewriting the equation to obtain

$$c_{jk} = \beta_{jk.S_{jk}} - \alpha_{jk} + \sum_{k+1 \leq l \leq j-1} \beta_{lk.S_{lk}} \alpha_{jl}, \quad V_k \in Pa_j. \quad (24)$$

We see that $c_{jk}$ is identifiable if each $\alpha_{ji}$, $i = k, \ldots, j-1$, is either zero or identifiable. And we have the following interesting result.

**Proposition 1** $c_{jk} = \beta_{jk.S_{jk}}$ if $\alpha_{ji} = 0$, $i = k, \ldots, j-1$.

The graphical condition for $\alpha_{ji}$ being zero is given in Lemma 2 in Section 5.1. Therefore we have a graphical condition under which $c_{jk}$ can be estimated by $\beta_{jk.S_{jk}}$.

To identify $\alpha_{ji}$'s we need to solve the set of equations in $E(NP_j)$ with $\alpha_{ji}$'s as unknowns, rewritten in the following

$$(V_k) : \beta_{jk.S_{jk}} = \alpha_{jk} - \sum_{k+1 \leq l \leq j-1} \beta_{lk.S_{lk}} \alpha_{jl}, \quad V_k \in NP_j. \quad (25)$$

In general we may have more equations than unknowns, or more unknowns than equations. These equations may not be linearly independent with each other. The difficulty of solving these linear equations lies in that the coefficients of these equations, the partial regression coefficients, are not independent free parameters. The partial regression coefficients are related to each other in a complicated way, and it is difficult to decide the rank of the set of linear equations since it is not easy to determine whether certain expressions of partial regression coefficients will cancel out each other and become identically zero.

To overcome this difficulty, it can be shown that a partial regression coefficient $\beta_{jk.S_{jk}}$ can be expressed in terms of the free parameters $c_{il}$'s and $\psi_{il}$'s along certain paths between $V_j$ and $V_k$ in the causal diagram. Then we are able to establish graphical conditions for the linear independence of the set of equations in $E(NP_j)$. First, we define some graphical notations.

### 5.1 Graph Background

A *path* between two nodes $X$ and $Y$ in a causal diagram consists of a sequence of consecutive edges of any type (directed or bidirected) (we will assume every variable appears only once in the path). A non-endpoint node $Z$ on a path is called a *collider* if two arrowheads on the path meet at $Z$, i.e. $\rightarrow Z \leftarrow, \leftrightarrow Z \leftrightarrow, \leftrightarrow Z \leftarrow, \rightarrow Z \leftrightarrow$; all other non-endpoint nodes on a path are *non-colliders*, i.e. $\leftarrow Z \rightarrow$, $\leftarrow Z \leftarrow, \rightarrow Z \rightarrow, \leftrightarrow Z \rightarrow, \leftarrow Z \leftrightarrow$. A *bidirected path* is a path such that every edge on the path is bidirected. The following concept of active path plays an important role in the analysis of the partial regression equations.

**Definition 1 (Active Path)** *A path between two nodes $X$ and $Y$ is said to be* active *given a set of nodes $Z$ if*

*(i) every non-collider on the path is not in $Z$, and*

*(ii) every collider on the path is in $Z$ or is an ancestor of a node in $Z$.*

The following lemma characterizes a class of active paths.

**Lemma 1** *For $k < j$, every node $V_l$ on an active path between $V_k$ and $V_j$ given $S_{jk}$ must be a collider, and $V_l < V_j$ (i.e., $V_l \in S_{jk}$). (see Figure 3)*

The following lemma gives a graphical condition for $\alpha_{jk}$'s being non-zero.

**Lemma 2** *For $k < j$, $\alpha_{jk} \not\equiv 0$ if there is a bidirected edge between $V_k$ and $V_j$ or there is a bidirected path between $V_k$ and $V_j$ such that every intermediate variable $V_l$ on the path is ordered ahead of $V_k$, $V_l < V_k$; otherwise $\alpha_{jk} = 0$.*

The following proposition, shown in [Pearl, 1998, Spirtes *et al.*, 1998], gives a graphical condition for $\beta_{jk.S_{jk}}$'s being non-zero.

**Proposition 2** $\beta_{jk.S_{jk}} = 0$ if there is no active paths between $V_j$ and $V_k$ given $S_{jk}$.

Proposition 2 essentially says that $\beta_{jk.S_{jk}} = 0$ if $V_j$ is conditionally independent of $V_k$ given $S_{jk}$. With Lemma 2 and Proposition 2, we are now able to write down the partial regression equations by inspecting the causal diagram.

### 5.2 Accessory Set

To identify $\alpha_{ji}$'s, we look for a set of variable $V_k$'s such that the set of equation $(V_k)$'s in (25) give a unique solutions to $\alpha_{ji}$'s. We will (informally) refer to these variables as accessory variables as they enable the identification of $\alpha_{ji}$'s. The idea is motivated by the widely used concept of instrumental variables and its extensions (auxiliary variables and instrumental set [Brito and Pearl, 2002b, Brito and Pearl, 2002a, Brito and Pearl, 2006]) which enable the identification of path coefficient $c_{ji}$'s. Next, we analyse a few examples to see conditions for a variable to serve as an accessory variable.

In the model in Figure 2, $Z$ serves as an instrumental variable for identifying the path coefficient $b$. In our framework, $Z$ also serves as an accessory variable for identifying $\alpha_{YX}$, which in turn leads to the identification of $b$, as shown by the following equations:

$$\beta_{YX} = b + \alpha_{YX} \quad (26)$$
$$\beta_{YZ.X} = -\beta_{XZ}\alpha_{YX}. \quad (27)$$

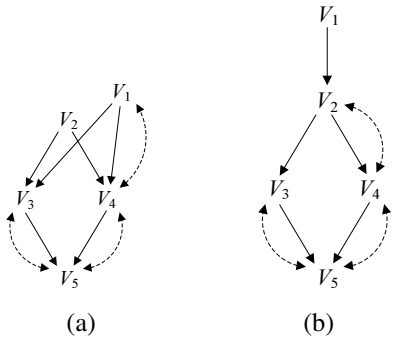

Figure 4: Examples of accessory variables.

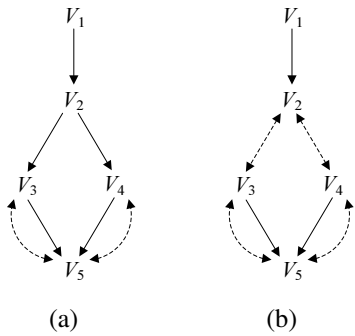

Figure 5: Examples of accessory variables.

In the model in Figure 1, $X$ is not qualified to be an instrumental variable for identifying path coefficient $c$. However $X$ can serve as an accessory variable for identifying $\alpha_{YW}$ (see Eq. (19)), which then leads to the identification of $c$ (see Eq. (18)).

The key observation seems to be that if there is an active path between $V_k$ and $V_j$ given $S_{jk}$ that ends with $V_i \leftrightarrow V_j$ for $k < i < j$, then $\alpha_{ji}$ will appear in the equation $(V_k)$. In general the equation $(V_k)$ may contain several $\alpha_{ji}$'s, then we need a set of equation $(V_k)$'s to simultaneously solve for $\alpha_{ji}$'s. For example, in the model in Figure 4(a), the equations $(V_1)$ and $(V_2)$ may be solved to identify $\alpha_{53}$ and $\alpha_{54}$ simultaneously:

$$(V_1) : \beta_{51.234} = -\beta_{31.2}\alpha_{53} - \beta_{41.23}\alpha_{54} \quad (28)$$
$$(V_2) : \beta_{52.134} = -\beta_{32.1}\alpha_{53} - \beta_{42.13}\alpha_{54}. \quad (29)$$

In the model in Figure 4(b), the equations $(V_1)$ and $(V_2)$ may also be solved to identify $\alpha_{53}$ and $\alpha_{54}$. In the model in Figure 5(a), $V_2$ may serve as an accessory variable for identifying $\alpha_{53}$ and $\alpha_{54}$ but $V_1$ can not since $\beta_{51.234} = 0$. Therefore we have one equation $(V_2)$ with two unknowns. In the model in Figure 5(b), both $V_1$ and $V_2$ can potentially serve as accessory variables, however we can show that the equations $(V_1)$ and $(V_2)$ are not independent.

The conditions for a set of variables $V_k$'s to serve as ac-cessory variables such that the set of equation $(V_k)$'s are linearly independent are characterized by the following formal definition of accessory variables.

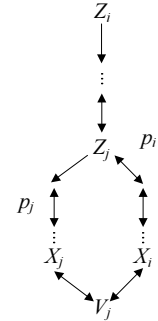

Figure 6: Definition of accessory set.

**Definition 2 (Accessory Set)** *A set of variables $Z = \{Z_1, \ldots, Z_k\} \subseteq NP_j$ is said to be an* accessory set *relative to $X = \{X_1, \ldots, X_k\} \subseteq Ne_j$ and $V_j$ if there exist a set of paths $p_1, \ldots, p_k$ such that*

1. *Either $Z_i = X_i$, $p_i$ degenerates into a node $X_i$; or $Z_i < X_i$, and $p_i$ is a path between $Z_i$ and $X_i$ such that every intermediate node $V_l$ is a collider on $p_i$ and $V_l < X_i$ (i.e., $p_i$ is an active path between $Z_i$ and $X_i$ given $S_{X_i Z_i}$); and*

2. *$p_i$ and $p_j$ do not share a common edge, and their only possible common variable must be either*

   - *$Z_j$, such that $p_i$ is a collider at $Z_j$ (unless $Z_j = X_i$, for this case, $p_i$ points at $X_i$), $Z_j \neq X_j$, and $p_j$ does not point to $Z_j$ (see Figure 6); or*
   - *$Z_i$, such that $p_j$ is a collider at $Z_i$ (unless $Z_i = X_j$, for this case, $p_j$ points at $X_j$), $Z_i \neq X_i$, and $p_i$ does not point to $Z_i$.*

For example, in Figure 4(a), $\{V_1, V_2\}$ is an accessory set relative to $\{V_3, V_4\}$ and $V_5$ with paths $V_1 \to V_3$ and $V_2 \to V_4$. In Figure 4(b), $\{V_1, V_2\}$ is an accessory set relative to $\{V_4, V_3\}$ and $V_5$ with paths $V_1 \to V_2 \leftrightarrow V_4$ and $V_2 \to V_3$ sharing a common variable $V_2$. In Figure 5(a), $\{V_1, V_2\}$ can not serve as an accessory set relative to $\{V_3, V_4\}$ while $\{V_2\}$ is an accessory set relative to $\{V_3\}$ (or $\{V_4\}$) and $V_5$. In Figure 5(b), $\{V_1, V_2\}$ can not serve as an accessory set relative to $\{V_3, V_4\}$ while $\{V_1\}$ or $\{V_2\}$ individually is an accessory set relative to $\{V_3\}$ (or $\{V_4\}$) and $V_5$.

**Theorem 1** *If a set of variables $Z = \{Z_1, \ldots, Z_k\} \subseteq NP_j$ is an accessory set relative to $X = \{X_1, \ldots, X_k\} \subseteq Ne_j$ and $V_j$, then the set of linear equations $E(Z) = \{(Z_1), \ldots, (Z_k)\}$ are linearly independent with respect to unknowns $\alpha_{V_j X_1}, \ldots, \alpha_{V_j X_k}$, that is, $E(Z)$ can be solved with respect to unknowns $\alpha_{V_j X_1}, \ldots, \alpha_{V_j X_k}$.*

An algorithm for finding an accessory set with maximum size is given in Section 6.

### 5.3 Solving Equations

Assume that we have identified an accessory set $Z = \{Z_1, \ldots, Z_k\}$. Let $U = \{\alpha_{V_j X_1}, \ldots, \alpha_{V_j X_m}\}$ be the set of unknowns appearing in the set of linear equations $E(Z)$. If $k = m$, then by Theorem 1, all the unknowns in $U$ are identified. In general, we may have $m \geq k$, that is, we have more unknowns than equations. One approach for solving the set of equations is to use Simon's causal ordering algorithm [Simon, 1953], extended in [Lu *et al.*, 2000]. Next we describe the algorithm.

We will say that a set of equations $S$ is *self-contained* if the number of equations is equal to the number of unknowns in $S$. A self-contained set is *minimal* if it does not contain any proper self-contained subsets. By Theorem 1, any self-contained subset $S$ of $E(Z)$ can be solved and all the unknowns in $S$ are identified.

To solve the set of linear equations in $E(Z)$, we start with *identifying* the mininal self-contained subsets in $E(Z)$. Then we *solve* these subsets, remove the equations in those subsets from $E(Z)$, and *substitute* the values of solved variables into remaining equations. The remaining set of equations is called the *derived system*. We keep applying identifying, solving, and substitution operations on derived system until either the derived system $D$ is empty, which happens if $E(Z)$ is self-contained, or there are no more self-contained subsets that can be identified, which means that the set of remaining variables in $D$ can not be solved.

After solving the set of equations $E(Z)$, the identifiability of the path coefficients $c_{jk}$'s can be determined by Eq. (24).

### 5.4 An Example

We illustrate the overall process of identifying path coefficients by an example. Consider the model in Figure 7. Assume that we want to identify the path coefficients associated with $V_7$ ($c_{74}$, $c_{75}$, and $c_{76}$). First we express the path coefficients in terms of $\alpha_{7i}$'s as follows

$$c_{76} = \beta_{76.12345} - \alpha_{76}. \tag{30}$$
$$c_{75} = \beta_{75.12346} - \alpha_{75}. \tag{31}$$
$$c_{74} = \beta_{74.12356} - \alpha_{74}. \tag{32}$$

Then we search for an accessory set using the algorithm given in Section 6. Assume that the algorithm returns $\{V_1, V_2\}$ as an accessory set. Then we attempt to solve the equations $(V_1)$ and $(V_2)$ given in the following:

$$(V_1) : \beta_{71.23456} = -\beta_{41.23}\alpha_{74} \tag{33}$$
$$(V_2) : \beta_{72.1345} = -\beta_{62.1345}\alpha_{76} - \beta_{52.134}\alpha_{75} - \beta_{42.13}\alpha_{74} \tag{34}$$

$(V_1)$ is self-contained and can be solved to identify $\alpha_{74}$ given by

$$\alpha_{74} = -\beta_{71.23456}/\beta_{41.23}. \tag{35}$$

We then substitute the value of $\alpha_{74}$ into the equation $(V_2)$, and we end up with one equation having two unknowns. Therefore we are not able to identify $\alpha_{75}$ and $\alpha_{76}$. In general, we can not make claims about being nonidentifiable since what we have presented is merely a sufficient criterion. In this particular model $\alpha_{75}$ and $\alpha_{76}$ are indeed nonidentifiable. Finally, by Eqs. (30)-(32), we conclude that $c_{74}$ is identified and is given by

$$c_{74} = \beta_{74.12356} + \beta_{71.23456}/\beta_{41.23}, \tag{36}$$

and $c_{75}$ and $c_{76}$ are both nonidentifiable. Note that equation $(V_3)$ as given below

$$(V_3) : \beta_{73.12456} = -\beta_{43.12}\alpha_{74}, \tag{37}$$

is not helpful for identifying $\alpha_{75}$ and $\alpha_{76}$, as expected. In fact $(V_3)$ leads to a constraint on the observed covariance matrix imposed by the model structure if we substitute the solved value of $\alpha_{74}$ into $(V_3)$ obtaining

$$\beta_{73.12456}\beta_{41.23} = \beta_{43.12}\beta_{71.23456}. \tag{38}$$

## 6 An Algorithm for Finding Accessory Set

In large models, it may not be easy to find an accessory set based on its definition. In this section, we provide an algorithm for finding an accessory set that contains the largest number of variables. We show that the problem can be reduced to a maximum flow problem.

A flow network $F = (V, E)$ is a directed graph in which each edge $(u, v) \in E$ has a nonnegative capacity $c(u, v) \geq 0$ (see, for example, [Cormen *et al.*, 1990]). We distinguish two vertices in a flow network: a source $s$ and a sink $t$. A flow in $F$ is a real-valued function $f : V \times V \to R$ that satisfies the capacity constraints $f(u, v) \leq c(u, v)$ and the flow conservation property (the amount of flow entering any vertex must be the same as the amount of flow leaving the vertex) among others. The value of a flow $f$ is defined as $|f| = \sum_{v \in V} f(s, v)$, that is, the total net flow out of the source. In the maximum flow problem, we are given a flow network $F$, with source $s$ and sink $t$, and we wish to find a flow of maximum value from $s$ to $t$.

To search for an accessory set relative to $V_j$, we construct a flow network $F$ as follows. The nodes of $F$ consists of:

- for every node $V_i < V_j$, add two nodes $V_i^-$ and $V_i^+$ into $F$.
- a source node $s$.
- a sink node $t$.

The edges of $F$ are:

- for every node $V_i < V_j$, add edge $V_i^- \to V_i^+$.
- for every edge $V_i \to V_l$, add edge $V_i^- \to V_l^+$.
- for every edge $V_i \leftrightarrow V_l$, add two edges $V_i^+ \to V_l^+$ and $V_i^+ \leftarrow V_l^+$.
- for every node $V_i \in Ne_j$ (those with $\alpha_{V_j X_i} \not\equiv 0$), add edge $V_i^+ \to t$.
- for every node $V_i \in NP_j$ (those with $c_{ji} = 0$), add $s \to V_i^-$.

We assign a capacity 1 to every edge in $F$. We also assign a node capacity of 1 to every node (except $s$ and $t$) in $F$ (this can be achieved by splitting every node into two and connecting them by an edge of capacity 1).

We then solve the maximum flow problem on the flow network $F$ (using, for example, Ford-Fulkerson algorithm). Since every edge has a capacity 1 and every node has a capacity 1, the computed flow $f$ represents a set of disjoint directed paths from $s$ to $t$. Let the set of paths be $q_i = s \to Z_i^- \to \ldots \to X_i^+ \to t, i = 1, \ldots, k$, where $k = |f|$. Each path $q_i$ in $F$ is interpreted as corresponding to a path $p_i$ from $Z_i$ to $X_i$ in the causal diagram $G$ as follows:

- An edge $V_a^- \to V_a^+$ corresponds to a node $V_a$.
- An edge $V_a^- \to V_b^+$ corresponds to an edge $V_a \to V_b$.
- An edge $V_a^+ \to V_b^+$ corresponds to an edge $V_a \leftrightarrow V_b$.

Note that if $Z_i = X_i$ then $p_i$ degenerates into a node $X_i$.

**Lemma 3** *The set of paths $p_1, \ldots, p_k$ satisfy the following properties*

- $Z_i \in NP_j$
- $X_i \in Ne_j$, that is, $\alpha_{V_j X_i} \not\equiv 0$.
- *Every intermediate node on $p_i$ is a collider.*
- *Property 2 in Definition 2.*

As an example, for the model shown in Figure 7(a), the flow network for searching an accessory set relative to $V_7$ is given in Figure 7(b), which shows a maximum flow solution. The flow corresponds to paths $V_1 \to V_2 \leftrightarrow V_4$ and $V_2 \to V_6$ in the causal diagram.

From Lemma 3, we see that the set of variables $Z = \{Z_1, \ldots, Z_k\}$ satisfy all the properties of an accessory set except that $Z_i$ or some intermediate nodes on the path $p_i$ may not be ordered ahead of $X_i$. Next we show that we can always obtain a set of variables $X \subseteq Ne_j$ such that $Z$ is an accessory set relative to $X$.

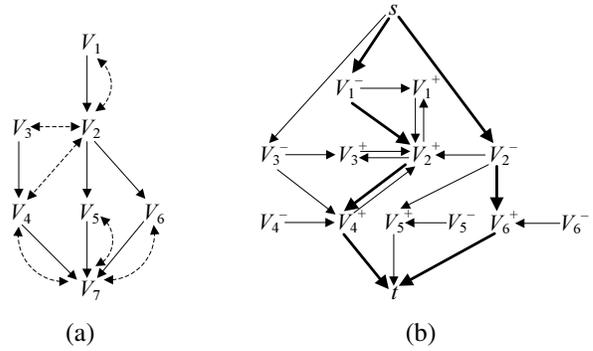

Figure 7: A SEM and corresponding flow network.

If there exist some nodes on $p_i$ that is ordered after $X_i$, let the variable ordered the last on $p_i$ be $X_{j_i}$.

**Lemma 4**   *1. $X_{j_i} \in Ne_j$, that is, $\alpha_{V_j X_{j_i}} \not\equiv 0$.*

*2. $X_{j_i}$ is not in $\{X_1, \ldots, X_k\}$.*

For every path $p_i$ that contains some nodes ordered after $X_i$, we replace $X_i$ with $X_{j_i}$ and the path $p_i$ with $p_i[Z_i, X_{j_i}]$, the subpath of $p_i$ between $Z_i$ and $X_{j_i}$. We have that every node on $p_i[Z_i, X_{j_i}]$ is ordered ahead of $X_{j_i}$.

**Theorem 2** *The set of variables $Z = \{Z_1, \ldots, Z_k\}$ is an accessory set relative to $\{X_{j_1}, \ldots, X_{j_k}\}$ and $V_j$, and $Z$ is an accessory set with maximum size.*

## 7 Conclusion and Discussion

The identification problem has been a long standing problem in the applications of linear SEMs. In this paper, we provide a procedure for identifying individual path coefficients, adding a new sufficient identification criterion for dealing with the problem in practice. Our method is based on the partial regression equations, which appears to be a new promising research direction towards solving the identification problem.

The closest related work is the instrumental set method in [Brito and Pearl, 2002a]. Both of the instrumental set and the accessory set method provide sufficient conditions for identification, and it is not clear whether one method has more identification power than the other. The advantage of the accessory set method presented in this paper is that we provide an algorithm for identifying an accessory set with maximum size, while it is not clear how we can find an instrumental set systematically, to say nothing of an instrumental set with maximum size.

In this paper, we have treated $\alpha_{ij}$'s as independent unknowns and obtained sufficient identification criterion. In certain class of models, $\alpha_{ij}$'s are indeed independent pa-rameters. We are investigating whether the criterion becomes also necessary in those situations.

**Acknowledgments**

This research was partly supported by NSF grant IIS-0347846.